\documentclass[prl,twocolumn,showpacs,preprintnumbers,amsmath,amssymb]{revtex4}

\usepackage{graphicx}
\usepackage{dcolumn}
\usepackage{bm}

\begin{document}


\title{X-Ray Scattering at Lanthanide $M_5$ Resonances: \\
Application to Magnetic Depth Profiling}

\author{H. Ott,$^{(1,\dag)}$ C. Sch{\"u}{\ss}ler-Langeheine,$^{(1,2)}$
E. Schierle,$^{(1)}$ A. Yu. Grigoriev,$^{(1,\dag\dag)}$
V. Leiner,$^{(3)}$, H. Zabel,$^{(3)}$ G. Kaindl,$^{(1)}$ and E. Weschke$^{(1,*)}$}
\affiliation{$^{(1)}$Institut f{\"u}r Experimentalphysik, Freie Universit{\"a}t
Berlin, D-14195 Berlin-Dahlem, Germany}
\affiliation{$^{(2)}$ II. Physikalisches Institut, Universit{\"a}t zu
K{\"o}ln,  D-50937 K{\"o}ln, Germany}
\affiliation{$^{(3)}$Institut f{\"u}r Experimentalphysik/Festk{\"o}rperphysik,
Ruhr-Universit{\"a}t Bochum, D-44780 Bochum, Germany}

\begin{abstract}
Quantitative analyses of x-ray scattering from thin films of Ho
and Dy metal at the $M_5$ resonances result in values of the
optical constants and the magnetic scattering lengths $f_{m}$,
with $f_{m}$ as large as $200 r_0$. The observation of first- and
second-order magnetic satellites allows to separate $f_{m}$ into
circular and linear dichroic contributions. This high magnetic
sensitivity, in conjunction with the tunable x-ray probing depth
across the resonance can be applied to monitor depth profiles of
complex magnetic structures, as e.g. of helical antiferromagnetic
domains in a Dy metal film.
\end{abstract}

\pacs{61.10.Eq,75.70.Ak,75.25.+z}

\date{\today}

\maketitle


Magnetism in thin films, nano\-structures, and other complex
materials is currently a field of considerable interest, where
diffraction and scattering methods can provide detailed insight
into spin structures and magnetic correlations. While magnetic
neutron diffraction had long been the method of choice, resonant
magnetic x-ray diffraction using synchrotron
radiation~\cite{gibbs:85} has emerged as a complementary
technique. Tuning the photon energy to an electronic core
excitation leads to an element-specific enhancement of magnetic
scattering~\cite{gibbs:88,hannon:88,isaacs:89,tonnerre:95,sacchi:98,lovesey:96}
that is particularly strong at the $L_{2,3}$ resonances of the
3$d$ transition elements and the $M_{4,5}$ resonances of
lanthanides and actinides. Quite early, Hannon {\em et al.}
predicted that at the $M_{4,5}$ resonances the magnetic
contribution $f_m$ to the scattering length $f$ should be of the
same order of magnitude as the contribution of charge scattering,
with values of $f_m$ up to $100 r_0$ ($r_0 =$ classical electron
radius)~\cite{hannon:88}. This agrees with the huge enhancement of
magnetic scattering by a factor of $\approx 10^7$ at the $M_4$
resonance of uranium~\cite{isaacs:89}, leading to sizeable
intensities along purely magnetic crystal truncation rods of
antiferromagnets, and permitting the study of surface magnetism of
UO$_2$~\cite{watson:00}. The magnetic sensitivity at resonance is
accompanied by strong x-ray absorption (XA), a fact that usually
renders a quantitative determination of $f_m$ from magnetic
superstructure peaks difficult~\cite{bernhoeft:98}. When used
properly, however, the strong XA can be exploited to vary the
probing depth of magnetic x-ray scattering.

In this Letter, we report on soft x-ray scattering from
lanthanide-metal films, providing a quantitative characterization
of magnetic scattering at the $M_{5}$ resonances, including a
separation of $f_m$ into circular and linear dichroic components
and a determination of the resonant index of refraction, $n =
1-\delta + i\beta$. With quantitative values for $\beta$ in the
resonance region, the x-ray probing depth can be tuned in a
controlled way, while retaining high magnetic sensitivity. The
probing depth can thus be varied independently of the scattering
vector, providing a tool for depth-resolved characterization of
complex magnetic structures, complementary to surface scattering
along the crystal truncation rods. As an example, a depth-resolved
study of the growth of helical antiferromagnetic (AFM) domains
across a magnetic phase transition in Dy metal is presented.

The helical AFM structures in Ho and Dy metal are well suited for
resonant magnetic soft x-ray studies, since the magnetic periods
match the x-ray wavelengths at the respective $M_5$ resonance. In
Ho metal this helical AFM structure persists in films down to 10
monolayers (ML)~\cite{weschke:04}, a thickness range where
absorption effects are reduced, simplifying a quantitative
determination of $f_m$. Soft x-ray studies were carried out in an
ultra-high-vacuum (UHV) ($\Theta / 2\Theta$) diffractometer on
{\em in-situ} grown lanthanide-metal films on
W(110)~\cite{weschke:04} and on Y/Ho/Y trilayer samples prepared
{\em ex situ} by molecular-beam epitaxy (MBE) on $a$-plane
sapphire~\cite{leiner:00}. For all samples, the $c$ axis was
perpendicular to the film surface. $M_5$ resonance data were taken
at beamline U49/1 of the Ber\-li\-ner
Elek\-tro\-nen\-spei\-cher\-ring f{\"u}r Synchrotronstrahlung
(BESSY) with linearly polarized x rays ($\pi$ polarization).
Resonant Dy $L_3$ data (at $\approx 8$ keV) were recorded at
beamline ID 10 A of the European Synchrotron Radiation Facility
(ESRF) in Grenoble.

\begin{figure}[t]
\includegraphics* [scale= 1.2] {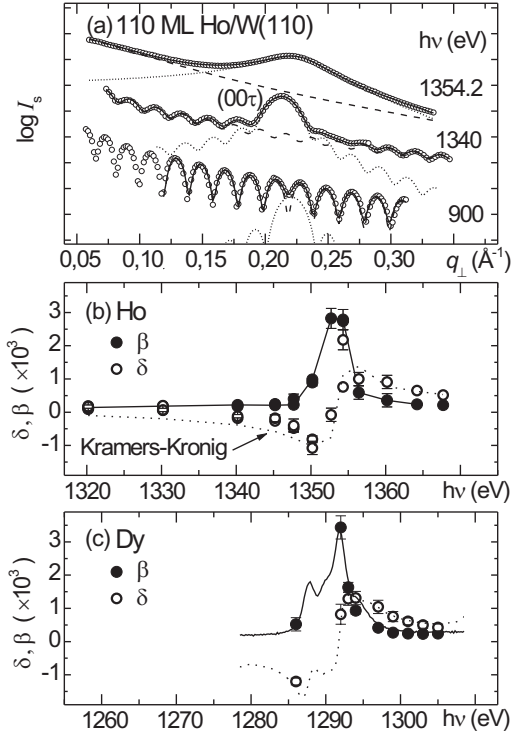}
\caption{(a) Specularly reflected intensities ($I_{s}$ on a
log-scale) from 110 ML Ho/W(110) versus momentum transfer
$q_\perp$ (data for different photon energies are vertically
offset). The solid lines represent fits with a superposition of
magnetic (dotted) and charge (dashed) contributions. Lower panels:
optical parameters $\beta$ (solid symbols) and $\delta$ (open
symbols) of (b) Ho and (c) Dy metal in the $M_5$ region. The
dotted lines represent the Kramers-Kronig transforms of the
$\beta$ values in (b), and $\beta$ obained from the scaled XA
spectrum (solid line) in panel (c).\label{fig:fig1}}
\end{figure}

The resonant enhancement of magnetic x-ray scattering at the $M_5$
resonance and the concomitant strong absorption are illustrated in
Fig.~1(a), which displays specular reflectivity curves from 110~ML
Ho on W(110) at 40~K, well below the bulk N{\'e}el temperature
($T_N = 131.2$~K). Far below resonance ($h\nu = 900$~eV), the
reflectivity is dominated by intensity oscillations (Kiessig
fringes) caused by interference of x rays scattered from the
surface with those scattered from the Ho/W
interface~\cite{als-nielsen:01}. At $h\nu = 1340$~eV, 14 eV below
the $M_5$ resonance maximum, the diffraction peak caused by the
magnetic superstructure, labeled $(00\tau)$, is already clearly
visible. At maximum resonance ($h\nu = 1354.2$~eV), the
substantial XA strongly alters the reflectivity curve: $(00\tau)$
is broadened due to the reduced number of layers that contribute
to the magnetic signal~\cite{bernhoeft:98}, and the Kiessig
fringes are completely suppressed, since the x rays no longer
penetrate to the Ho/W interface.

As a first step towards a determination of $f_m$, the resonant
optical constants were derived from reflectivity curves by fit
analyses using the Parratt formalism~\cite{parratt:54}. Far from
resonance, tabulated values of $\beta$ and
$\delta$~\cite{henke:93} are reliable and were used to fit the
900-eV data, yielding the structural parameters of the
film~\cite{structuralparameter}. The optical constants at a given
resonance energy were then obtained from a fit of the respective
reflectivity curve with $\beta$ and $\delta$ as the only
adjustable parameters. The superimposed magnetic peak was
described by the structure factor of a helix, taking a mean
magnetic roughness and an angle-dependent polarization factor into
account~\cite{lovesey:96}. Refraction and absorption corrections
to the shape of the magnetic peak were accounted for by a complex
scattering vector.

The resulting $\beta$ and $\delta$ values are plotted in
Fig.~1(b); they are consistent with a Kramers-Kronig (KK)
transform of $\beta$ that reproduces $\delta$ rather well. At
maximum resonance, $\beta = (2.8 \pm 0.3)\cdot 10^{-3}$ is
obtained, a value that fits well with recent results for Gd and
Tb~\cite{prieto:03}, and corresponds to a photon attenuation
length of $1/\mu=\lambda/4\pi\beta\approx 260$~{\AA}, i.e., an
effective probing depth of only $\approx 28$~{\AA} at the magnetic
peak position (scattering angle $\Theta \approx 12.5^\circ$ with
respect to the sample surface). Similar data were obtained for Dy
as shown in Fig.~1(c). Here, the solid line represents $\beta$
values obtained from the Dy $M_5$ XA spectrum, the KK transform of
which consistently reproduces the $\delta$ values obtained from
the reflectivity curves.

For a quantitative determination of $f_m$, a thin Y/Ho/Y film was
used that exhibits a larger $\tau$ compared to
Ho/W(110)~\cite{leiner:00} and hence a smaller charge-scattering
background at ($00\tau$). Figure 2(a) displays the $M_5$ XA
spectrum of Ho with its atomic $3d^{9}4f^{11}$ final-state
multiplet that separates into three subspectra with $\Delta J =0,
\pm 1$~\cite{goedkoop:88,schille:93}. In the presence of magnetic
order, the $J$ states split into $M_J$ sublevels, and the
transitions are governed by the selection rule $\Delta M_J =0, \pm
1$. In the heavy lanthanides, the transition probabilities for a
given $\Delta J$ are dominated by a {\em single} $\Delta M_J$
value, and the subspectra can then be identified approximately by
$\Delta M_J =0, \mp 1$, respectively~\cite{schille:93}. These
dipole transitions determine the resonant scattering
length~\cite{hannon:88,lovesey:96}
\begin{equation}\label{eq1}
f = ({\bf e'}\cdot {\bf e})\cdot f_0 - i\left( {\bf e'}\times {\bf
e}\right) \cdot {\bf m}\cdot f_m^{c} + \left( {\bf e'} \cdot {\bf
m}\right) \left( {\bf e}\cdot {\bf m}\right)\cdot f_m^{l} ,
\end{equation} \noindent with $f_0 =
a\left(F_{+1}^{1}+F_{-1}^{1}\right)$, $f_m^{c} = a \left(
F_{+1}^{1}-F_{-1}^{1}\right)$, and $f_m^{l} = a
\left(2F_0^1-F_{+1}^{1}-F_{-1}^{1}\right)$. Here, the $F_{\Delta
M_J}^1$ denote the energy-dependent dipole oscillator strengths
with $\Delta M_J = 0, \pm 1$, and $\bf e$ and $\bf e'$ the
polarization vectors of incident and scattered x rays,
respectively. $a=(3/4\pi k)$ is a wave-vector dependent factor and
$\bf m$ is the unit vector in direction of the local magnetic
moment.

\begin{figure}[t]
\includegraphics* [scale= 1.2]{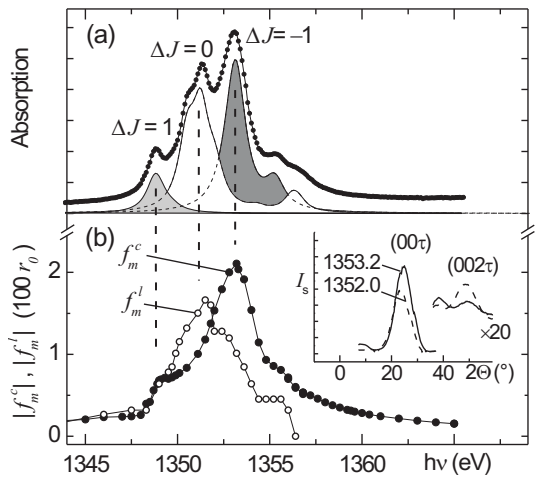}
\caption{(a) $M_5$ XA spectrum of 31 ML Ho/W(110) recorded via
sample drain current. The subspectra represent calculated $\Delta
J = 0,\pm1 $ transitions~\cite{goedkoop:88}. (b) Magnetic
scattering lengths $|f_m^{c}|$ and $|f_m^{l}|$. The inset shows
reflectivity curves (on a linear scale) from a 16-ML Ho MBE film,
taken at the two given photon energies (in eV). \label{fig:fig2}}
\end{figure}

For Ho, circular ($f_m^{c}$) and linear ($f_m^{l}$) dichroic terms
in Eq.~1 are readily distinguished, since the corresponding
diffraction peaks at $(00\tau)$ and $(002\tau)$ are well separated
in momentum space (inset in Fig.~2(b)). The resonant $(00\tau)$
peak at $2\Theta \approx 25^\circ$ is due to $f_m^{c}$ with a
polarization factor {\em linear} in $\bf m$. Characterized by a
sinusoidal modulation, the magnetic structure of Ho contains no
higher harmonics, and the $(002\tau)$ peak at $2\Theta \approx
50^\circ$ is thus solely a resonance effect due to
$f_m^{l}$~\cite{mannix:00}. With a polarization factor {\em
quadratic} in $\bf m$, $f_m^{l}$ gives rise to a resonant peak at
$(002\tau)$, since ${\bf m}^2$ oscillates with half the magnetic
period. The linear dichroic term $f_m^{l}$ generally probes
elements that preserve time-reversal symmetry~\cite{wilkins:04};
therefore it can be used to study the ordering of quadrupole
moments in non-spherical charge densities. In the present case of
AFM Ho metal, however, a distinction of quadrupolar and spin
linear dichroism is of no significance, since the strong
spin-orbit interaction of the atomic-like $4f$ states couples the
arrangement of the quadrupole moments to the spin
structure~\cite{amara:98}. The $(00\tau)$ and $(002\tau)$ can be
entirely described by $f_m^{c}$ and $f_m^{l}$, respectively,
according to the $\Delta M_J = 0, \pm 1 $ transitions from the
$M_J$ sublevels of the helical AFM ground state of Ho.

From the respective integrated intensities, $|f_m^{c}|$ and
$|f_m^{l}|$ were calculated, taking polarization
factors~\cite{lovesey:96} and absorption corrections into account
($\beta$ values from Fig.~1(b)). The resulting $|f_m^{c}|$ and
$|f_m^{l}|$ are plotted in Fig.~2(b), revealing a clear
correspondence to the subspectra in Fig.~2(a): $|f_m^{c}|$ peaks
at the maxima of the $\Delta J = \pm 1$ subspectra, whereas
$|f_m^{l}|$ peaks at the maximum of the $\Delta J = 0$
subspectrum. Thus, the $3d^{9}4f^{11}$ multiplet of Ho allows to
identify the circular and linear dichroic contributions to the
scattering length; such a separation is not equally
straightforward in case of the actinides~\cite{mannix:00}. We
obtain $|f_m^{c}| = 200 r_0$ and $|f_m^{l}| = 160 r_0$ at the
respective resonance maxima, i.e. values that are somewhat larger,
but of the same order of magnitude as predicted~\cite{hannon:88}.

\begin{figure}[b]
\includegraphics* [scale= 1.2] {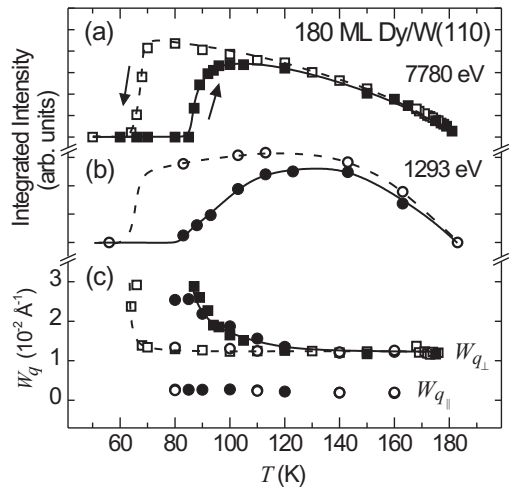}
\caption{Magnetic diffraction data characterizing the
FM/helical-AFM phase transition in 180-ML Dy/W(110) upon cooling
down (open symbols) and warming up (filled symbols); lines serve
as guides to the eye. (a) Integrated intensities of (002-$\tau$)
at the $L_3$ resonance (7780~eV). (b) Analogous data from
(00$\tau$) recorded at the $M_5$ resonance. (c) Widths of magnetic
peaks in the direction parallel ($W_{q_{\|}}$) and perpendicular
($W_{q_{\perp}}$) to the film plane recorded at 7780~eV (squares)
and at 1305~eV (circles).\label{fig:fig3}}
\end{figure}

An application that exploits both the tunable x-ray probing depth
and the high magnetic sensitivity across the resonance is the
study of depth-dependent inhomogeneous magnetic structures, even
in case of chemically homogeneous materials. We demonstrate the
potential of the method for the growth of helical AFM domains at
the ferromagnetic (FM)/helical-AFM first-order phase transition of
Dy metal~\cite{wilkinson:61}. As shown in Fig.~3(a) for a 180-ML
film of Dy on W(110), this transition exhibits substantial
hysteresis. Here, the integrated intensity of (002-$\tau$) is
displayed~\cite{gibbs:85,gibbs:88}, recorded at the $L_3$
resonance of Dy (h$\nu = 7780$~eV), where all layers of the film
contribute about equally to the magnetic signal; the x-ray probing
depth of $\approx 9$~$\mu$m is much larger than the film thickness
of $\approx 500$~\AA. Due to the weaker magnetic sensitivity at
$L_3$ as compared to $M_5$, a polarization analysis was required
to separate (002-$\tau$) from the charge-scattering
background~\cite{gibbs:85,gibbs:88}. When cooling to the FM phase
(open symbols), (002-$\tau$) disappears at $\approx 70$~K; it
reappears when heating, but with a delay of $\approx 20$~K (solid
symbols). Notably, the (002-$\tau$) intensity does not recover
abruptly, but remains below the cooling-down curve up to $\approx
140$~K, indicating a temperature-dependent growth of helical AFM
domains. This is further characterized by the widths of the
magnetic diffraction peaks displayed in Fig.~3(c), both in the
direction perpendicular ($W_{q_{\perp}}$) and parallel
($W_{q_{\|}}$) to the film plane. $W_{q_{\perp}}$ was determined
both for (002-$\tau$) at the $L_3$ (squares) and for (00$\tau$) at
the $M_5$ (circles) resonance. The smaller $W_{q_{\|}}$ was
measured only at the $M_5$ resonance, where sufficient momentum
resolution could be achieved. When cooling, $W_{q_{\perp}}$
remains essentially constant down to $\approx 70$~K;
$W_{q_{\perp}}\approx 1.2\times 10^{-2}$~\AA$^{-1}$ corresponds to
180 ML of Dy, i.e., the helical AFM structure extends through the
whole film. Below $\approx 70$~K, $W_{q_{\perp}}$ increases
abruptly with the decay of the helical AFM order and -- with
increasing temperature -- exhibits the same hysteresis as the
intensity of (002-$\tau$). In contrast, $W_{q_{\|}}$ is
essentially constant in the studied temperature range, showing
that the helical AFM domains develop in a laterally homogeneous
way in the direction perpendicular to the film plane.

\begin{figure}[t]
\includegraphics* [scale= 0.54] {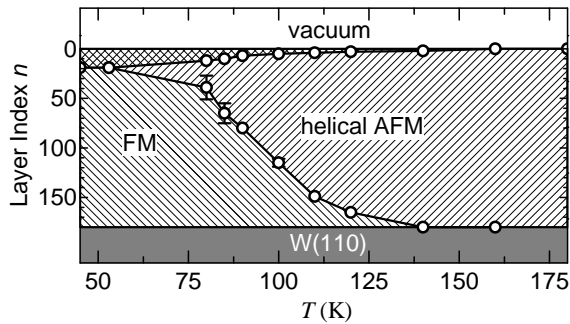}
\caption{Temperature-dependent growth of helical AFM domains in
180-ML Dy/W(110) in the direction perpendicular to the film plane.
For details, see text. \label{fig:fig4}}
\end{figure}

For a depth-resolved characterization of domain growth, ($00\tau$)
was studied at the Dy $M_5$ resonance. Fig.~3(b) displays the
integrated intensities obtained at h$\nu = 1293$~eV where the
effective x-ray probing depth is only $\approx 35$~{\AA}. Here,
the delayed formation of the helical AFM phase upon heating is
more pronounced than at 7780~eV excluding a nucleation of the
helical domain in the {\em topmost} surface layer. Instead, the
domain nucleates close to the surface as discussed below. A
further detail of the reflectivity curves recorded at highest
surface sensitivity (data not shown here) is the occurrence of a
magnetic signal at smaller scattering angles in addition to
($00\tau$). This is the signature of a surface AFM structure with
a larger mean modulation period in the surface layers of the film
that develops when cooling below $\approx 125$~K and that
disappears upon heating above this temperature. Such a structure
had previously been reported for a 10-ML thick Ho
film~\cite{weschke:04b}, and results from the tendency of helical
AFM lanthanide films to favor FM order in the surface
region~\cite{weschke:04}.

The complete depth profile, shown in Fig.~4, was derived from data
recorded with various x-ray probing depths across the Dy $M_5$
resonance. The $\beta$ values determined for the same film
(Fig.~1(c)) allowed an analysis, where the magnetic scattering
amplitude of an individual layer $n$ at depth $d_n$ is reduced by
$e^{-\mu d_n/sin\theta}$~\cite{bernhoeft:98}. The consistent
analysis of temperature-dependent intensities leads to the
following scenario: Upon cooling, the whole film orders helical
AFM below $T_N \approx 179$~K. Below $\approx 125$~K, the surface
AFM structure develops as described above, consistent with the
minute increase of $W_{q_{\perp}}$ by $\approx 1$\% in this
temperature region. Below $\approx 70$~K, the film turns FM,
except for the topmost $\approx 19$ layers that retain the surface
AFM structure (cross-hatched area). Upon heating, the helical AFM
domain nucleates below this surface region and grows with
increasing $T$ towards the two interfaces. When reaching $\approx
140$~K, it has developed across the whole film.

We point out that the present approach of exploiting the tunable
x-ray probing depth across a resonance is not restricted to
magnetic signals, and is also applicable to systems containing
$3d$ transition elements, where at the $L_{2,3}$ resonances
similarly strong absorption has been observed~\cite{kortright:00}.
Such depth-dependent diffraction studies will be particularly
interesting with focused x-ray beams providing additionally
lateral resolution~\cite{evans:02}.

Expert support by the staff members of BESSY and the ESRF is
gratefully acknowledged, particularly the commitment of G.
Gr\"ubel. We acknowledge helpful discussions with M. W. Haverkort.
The work was financially supported by the BMBF, projects
05KS1KEE/8 and 03ZA6BC2, the Sfb-290 (TPA06) of the DFG, and the
Landesministerium NRW f\"ur Wissenschaft und Forschung.


\begin{thebibliography}{widest-label}

\bibitem[$^*$] {ew} Corresponding author: \\
eugen.weschke@physik.fu-berlin.de

\bibitem[$^\dag$] {ag} present address:
II. Physikalisches Institut, Universit{\"a}t zu K{\"o}ln,  D-50937
K{\"o}ln, Germany

\bibitem[$^{\dag\dag}$] {vl} present address:
Department of Materials Science and Engineering, University of
Wisconsin, Madison, WI 53706, U.S.A.

\bibitem{gibbs:85}
D. Gibbs {\em et al.},
Phys. Rev. Lett. {\bf 55}, 234 (1985).

\bibitem{gibbs:88}
D. Gibbs {\em et al.},
Phys. Rev. Lett. {\bf 61}, 1241 (1988).

\bibitem{hannon:88}
J.~P. Hannon{\em et al.},
Phys. Rev. Lett. {\bf 61},  1245 (1988).

\bibitem{isaacs:89}
E. D. Isaacs {\em et al.},
Phys. Rev. Lett. {\bf 62}, 1671 (1989).

\bibitem{tonnerre:95}
J. M. Tonnerre {\em et al.},
Phys. Rev. Lett. {\bf 75}, 740 (1995).

\bibitem{sacchi:98}
M. Sacchi {\em et al.},
Phys. Rev. Lett. {\bf 81}, 1521 (1998).

\bibitem{lovesey:96}
S. W. Lovesey and S. P. Collins, {\em X-ray Scattering and Absorption by Magnetic Materials},
Clarendon Press, Oxford (1996), p. 180ff.

\bibitem{watson:00}
G. M. Watson {\em et al.},
Phys. Rev. B {\bf 61}, 8966 (2000).

\bibitem{bernhoeft:98}
N. Bernhoeft {\em et al.},
Phys. Rev. Lett. {\bf 81}, 3419 (1998).

\bibitem{weschke:04}
E. Weschke, {\em et al.},
Phys. Rev. Lett. {\bf 93}, 157204 (2004).

\bibitem{leiner:00}
V. Leiner, {\em et al.},
Physica B {\bf 283}, 167 (2000).

\bibitem{als-nielsen:01}
J. Als-Nielsen and D. McMorrow, {\em Elements of Modern X-Ray Physics}, John Wiley \& Sons,
New York (2001).

\bibitem{parratt:54}
L.~G. Parratt, Phys. Rev. {\bf 95}, 359 (1954).

\bibitem{henke:93}
B.~L. Henke, E.~M. Gullikson, and J.~C. Davis,
At. Data Nucl. Data Tables {\bf 54}, 181 (1993);
http://www-cxro.lbl.gov/optical\_constants.

\bibitem{structuralparameter}
For the present film, we obtain a thickness of 110 ML [($309 \pm 1$)~{\AA}],
and a combined surface and interface
roughness of ($1.0 \pm 0.5$)~\AA.

\bibitem{prieto:03}
J. E. Prieto {\em et al.},
Phys. Rev. B {\bf 68}, 134453 (2003).

\bibitem{goedkoop:88}
J. B. Goedkoop {\em et al.},
Phys. Rev. B {\bf 37}, 2086 (1988).

\bibitem{schille:93}
J. Ph. Schill{\'e} {\em et al.},
Phys. Rev. B {\bf 48}, 9491 (1988).

\bibitem{mannix:00}
D. Mannix {\em et al.},
Phys. Rev. B {\bf 62}, 3801 (2000).

\bibitem{wilkins:04}
S. B. Wilkins {\em et al.},
Phys. Rev. B {\bf 70}, 214402 (2004).

\bibitem{amara:98}
M. Amara and P. Morin
J. Phys.: Condens. Matter {\bf 10}, 9875 (1998).

\bibitem{wilkinson:61}
M. K. Wilkinson {\em et al.},
J. Appl. Phys. 32, 48S (1961).

\bibitem{weschke:04b}
E. Weschke {\em et al.}, Synchr. Rad. News {\bf 295}, 1042 (2002).

\bibitem{kortright:00}
J. B. Kortright and Sang-Koog Kim, Phys. Rev. B {\bf 62}, 12216 (2000).

\bibitem{evans:02}
P. G. Evans {\em et al.}, Science {\bf 295}, 1042 (2002).

\end{thebibliography}
\end{document}